\documentclass[lettersize,journal]{IEEEtran}
\usepackage{amsmath,amsfonts}
\usepackage{algorithmic}
\usepackage{array}
\usepackage{subfigure}
\usepackage{textcomp}
\usepackage{stfloats}
\usepackage{url}
\usepackage{verbatim}
\usepackage{graphicx}
\usepackage{cite}

\usepackage{mathrsfs}
\usepackage{amssymb}
\usepackage{multirow}
\usepackage[ruled]{algorithm2e}
\usepackage{float}
\usepackage{xcolor}
\usepackage[colorlinks=true,linkcolor=black,citecolor=blue,urlcolor=blue,]{hyperref}
\begin{document}

\title{SComCP: Task-Oriented Semantic Communication for Collaborative Perception}

\author{

	Jipeng~Gan,~\IEEEmembership{Student Member,~IEEE,} 
  Yucheng~Sheng,~\IEEEmembership{Student Member,~IEEE,}
  Hua~Zhang,~\IEEEmembership{Member,~IEEE,}
	Le~Liang,~\IEEEmembership{Member,~IEEE,}
  Hao~Ye,~\IEEEmembership{Member,~IEEE,}
  Chongtao Guo,~\IEEEmembership{Member,~IEEE,}
	and~Shi~Jin,~\IEEEmembership{Fellow,~IEEE}

  \thanks{Jipeng~Gan, Yucheng~Sheng, Hua~Zhang, Le Liang and Shi Jin are with the National Mobile Communications Research Laboratory, Southeast University, Nanjing 210096, China (e-mail: \{jpgan, shengyucheng, huazhang, lliang, jinshi\}@seu.edu.cn). Le Liang is also with the Purple Mountain Laboratories, Nanjing 211111, China.

  Hao Ye is with the Department of Electrical and Computer Engineering, University of California, Santa Cruz, CA 95064, USA (e-mail: yehao@ucsc.edu).

  Chongtao Guo is with the College of Electronics and Information
     Engineering, Shenzhen University, Shenzhen 518060, China (e-mail: ctguo@
     szu.edu.cn).
}
  
  }

\maketitle

\begin{abstract}
  Reliable detection of surrounding objects is critical for the safe operation of connected automated vehicles (CAVs). However, inherent limitations such as the restricted perception range and occlusion effects compromise the reliability of single-vehicle perception systems in complex traffic environments. 
  Collaborative perception has emerged as a promising approach by fusing sensor data from surrounding CAVs with diverse viewpoints, thereby improving environmental awareness. Although collaborative perception holds great promise, its performance is bottlenecked by wireless communication constraints, as unreliable and bandwidth-limited channels hinder the transmission of sensor data necessary for real-time perception.
  To address these challenges, this paper proposes SComCP, a novel task-oriented semantic communication framework for collaborative perception. Specifically, SComCP integrates an importance-aware feature selection network that selects and transmits semantic features most relevant to the perception task, significantly reducing communication overhead without sacrificing accuracy.
  Furthermore, we design a semantic codec network based on a joint source and channel coding (JSCC) architecture, which enables bidirectional transformation between semantic features and noise-tolerant channel symbols, thereby ensuring stable perception under adverse wireless conditions.
  Extensive experiments demonstrate the effectiveness of the proposed framework. In particular, compared to existing approaches, SComCP can maintain superior perception performance across various channel conditions, especially in low signal-to-noise ratio (SNR) scenarios. In addition, SComCP exhibits strong generalization capability, enabling the framework to maintain high performance across diverse channel conditions, even when trained with a specific channel model.

\end{abstract}

\begin{IEEEkeywords}
	Semantic communication, autonomous driving, collaborative perception, vehicle-to-vehicle communication, 3D object detection.
\end{IEEEkeywords}

\section{Introduction}	
	
\begin{figure}[t!]
  \centering
  \includegraphics[width=3.5in]{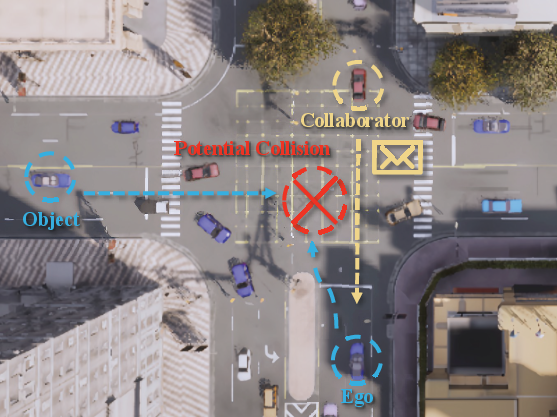}
  \caption{An illustrative example of collaborative perception scenario.}
  \label{fig1}
\end{figure}

\IEEEPARstart{W}{ith} the rapid advancement of artificial intelligence and wireless communication technologies, autonomous driving is emerging as a transformative force in intelligent transportation systems \cite{AV}.
Achieving Level 4-5 autonomy necessitates precise perception of complex traffic environments, which is crucial for downstream tasks such as object tracking, motion prediction, and navigation \cite{l5,l51}. 
Traditional vehicular environment perception relies solely on onboard sensors such as cameras, LiDAR, and mmWave radar, among which LiDAR has emerged as a mainstream solution due to its superior three-dimensional positioning capability \cite{sensors, lidar}. However, single-vehicle perception systems suffer from limitations such as line-of-sight obstructions, homogeneous viewpoints, and restricted perception range, which may compromise road safety and traffic efficiency \cite{single}. 
To overcome these limitations, collaborative perception has emerged as a promising paradigm that leverages vehicle-to-vehicle (V2V) communication to share perception information among connected automated vehicles (CAVs), thereby enhancing overall perception performance \cite{CP1,CP2,CP3,  av9}. 
As shown in Fig.~\ref{fig1}, the ego vehicle can utilize shared perception information from collaborating CAVs to detect occluded objects, enabling preemptive control actions to avoid potential collisions.

Collaborative perception can be categorized into three types—early collaboration, intermediate collaboration, and late collaboration—based on the stage at which information is integrated \cite{fusion}.
Early collaboration enables CAVs to fuse raw perception data (e.g., LiDAR point clouds) from both their own sensors and those received from other collaborators. This approach maximizes perception performance but requires substantial communication resources due to the high data volume \cite{early1,early2}.
In contrast, late collaboration requires CAVs to exchange fully processed perception information (e.g., object detection results), enabling the fusion of object-level data \cite{late1,late2}. While this approach significantly reduces communication overhead, it limits perception performance due to the loss of rich contextual information.
To reach a compromise between perception performance and communication overhead, intermediate collaboration allows CAVs to transmit intermediate features (e.g., semantic information) extracted from raw sensor data, thereby reducing data size while preserving critical scene context \cite{inter1,inter2, fusion5}. As a result, intermediate fusion has become a mainstream approach in current collaborative perception methods, especially when communication resources are constrained.

Building on the intermediate fusion paradigm, many recent works have sought to balance the perception performance and communication overhead \cite{tradeoff1, tradeoff2}. For instance, \cite{when2comm} proposed a communication framework for selecting the most relevant CAVs to cooperate in a distributed manner. Compared to traditional broadcasting methods, this framework reduces transmission load by decoupling the communication phase, which includes building communication groups and determining when to communicate. 
Additionally, a spatial perception graph neural network is employed to effectively aggregate LiDAR feature information from proximate CAVs, enabling an end-to-end learning-based source coding method in \cite{v2vnet}. Similarly, \cite{1d} utilized one-dimensional convolutional neural networks (CNNs) to compress perception features before transmitting them to collaborators. 
It is worth noting that the aforementioned feature compression-based studies generally assume that cooperating CAVs must share perception information regarding all spatial regions. However, this assumption overlooks the fact that many regions may contain task-irrelevant semantic information, leading to redundant transmission and unnecessary bandwidth consumption. To mitigate this inefficiency, \cite{where2comm} proposed a communication strategy based on a spatial confidence map, which significantly reduces the communication load by conveying only relevant semantic information.
Nevertheless, this method heavily relies on the accuracy of the confidence map, and its local feature extraction limits the ability to capture global spatial context, leading to potential overlook of critical perception information. Moreover, most existing methods adopt fixed transmission schemes that do not adapt to variations in semantic information richness across different traffic environments, thereby constraining overall transmission efficiency.

A fundamental limitation of most existing collaborative perception approaches is the assumption of an ideal V2V communication channel, disregarding real-world wireless channel impairments such as fading, noise interference, and bandwidth constraints.
This neglect of realistic channel conditions can undermine the reliability of collaborative perception, thereby significantly degrading perception accuracy and potentially compromising the safety of autonomous driving systems in real-world deployments \cite{safety, challenges1}.
While traditional communication schemes based on separated source–channel coding (SSCC) provide a theoretical framework for error control, they are inherently limited by the cliff effect in low signal-to-noise ratio (SNR) scenarios \cite{channel, cliff}. Specifically, when the channel quality degrades to a level where error-free decoding cannot be guaranteed, traditional communication schemes suffer an abrupt performance collapse due to decoding failures, rendering them less competitive in bandwidth-constrained V2V networks \cite{bandwidth}.
Fortunately, semantic communication offers a promising solution by extracting and transmitting task-oriented semantic information, focusing on conveying intended meaning rather than precise bits, thereby reducing communication load and ensuring the reliable transfer of critical semantic information \cite{semantic}. Specifically, semantic communication based on deep joint source-channel coding (JSCC) architectures has demonstrated excellent performance in multimodal transmission tasks (e.g., LiDAR \cite{jscc_lidar}, images \cite{jscc_image}, and speech \cite{jscc_speech}) under dynamic channel conditions \cite{jscc}.
\textcolor{black}{Recent advancements in semantic communication, such as the framework introduced by \cite{franklin}, have demonstrated significant potential for collaborative perception. By employing a CNN-based JSCC codec network to transmit relevant perception features, the method achieve graceful performance degradation under noisy channel conditions, surpassing traditional separation-based schemes. However, the stringent safety requirements of autonomous driving demand greater robustness than existing frameworks provide, particularly under severe channel conditions where communication failures could prove catastrophic.}

\textcolor{black}{
A fundamental challenge in collaborative perception is ensuring both communication efficiency and transmission reliability in real-world vehicular networks. Existing methods exhibit two primary limitations: scene-agnostic transmission schemes that lead to excessive bandwidth consumption due to redundant data exchange, and unreliable communication that distorts perception features under low SNR conditions, thereby compromising vehicular safety. These interconnected challenges underscore the urgent need for a more holistic and adaptive communication framework.
In this paper, we propose a task-oriented semantic communication framework to address these limitations. The proposed approach is guided by two core principles. First, to achieve communication efficiency, the transmission strategy is dynamically adapted to the semantic context of the environment, ensuring that communication resources are prioritized for the most critical information relevant to the perception task. Second, to ensure reliability, semantic codec networks are designed to generate robust feature representations, maintaining high perception accuracy even under adverse channel conditions. By jointly optimizing both the information to be transmitted and its protection mechanisms, the proposed framework enables reliable and efficient collaborative perception in dynamic and challenging wireless environments.}
The main contributions of this work are summarized as follows:

	\begin{itemize}
		\item[$\bullet$] We propose SComCP, a novel task-oriented semantic communication framework for collaborative perception. The framework unifies feature extraction and selection, semantic encoding, transmission, reconstruction and fusion, and object detection into a single end-to-end pipeline. SComCP significantly reduces bandwidth consumption by transmitting only task-relevant semantic information, while effectively mitigating the effects of channel impairments to maintain high perception performance.
		\item[$\bullet$]We design an importance-aware feature selection network that learns to identify and prioritize semantically critical perception features. The network dynamically adjusts the transmission volume according to the semantic information richness of the scene, enabling adaptive data exchange to optimize both communication efficiency and perception performance in dynamic V2V environments.
		\item[$\bullet$] We develop a deep JSCC-based semantic codec network that generates robust feature representations, significantly enhancing transmission reliability under low SNR conditions. Notably, the codec exhibits strong generalization across diverse wireless channel models, despite being trained on a single model, thereby reducing the need for retraining in complex real-world environments.
		\item[$\bullet$] We conduct extensive experiments and ablation studies to validate the effectiveness of SComCP. Quantitative results demonstrate that our framework achieves superior perception accuracy and transmission efficiency compared to state-of-the-art methods. The ablation study further confirms the individual contributions of the proposed components to the overall system performance.
	\end{itemize}

  The remainder of this paper is organized as follows. Section II introduces the system model and problem formulation. In Section III, we present the ScomCP framework in detail, including the proposed importance-aware feature selection module, semantic codec network, loss function design, and training strategy. Section IV provides simulation results, followed by the conclusion in Section V.

  \begin{figure*}[htbp]
  \centering
  \includegraphics[width=0.93\linewidth]{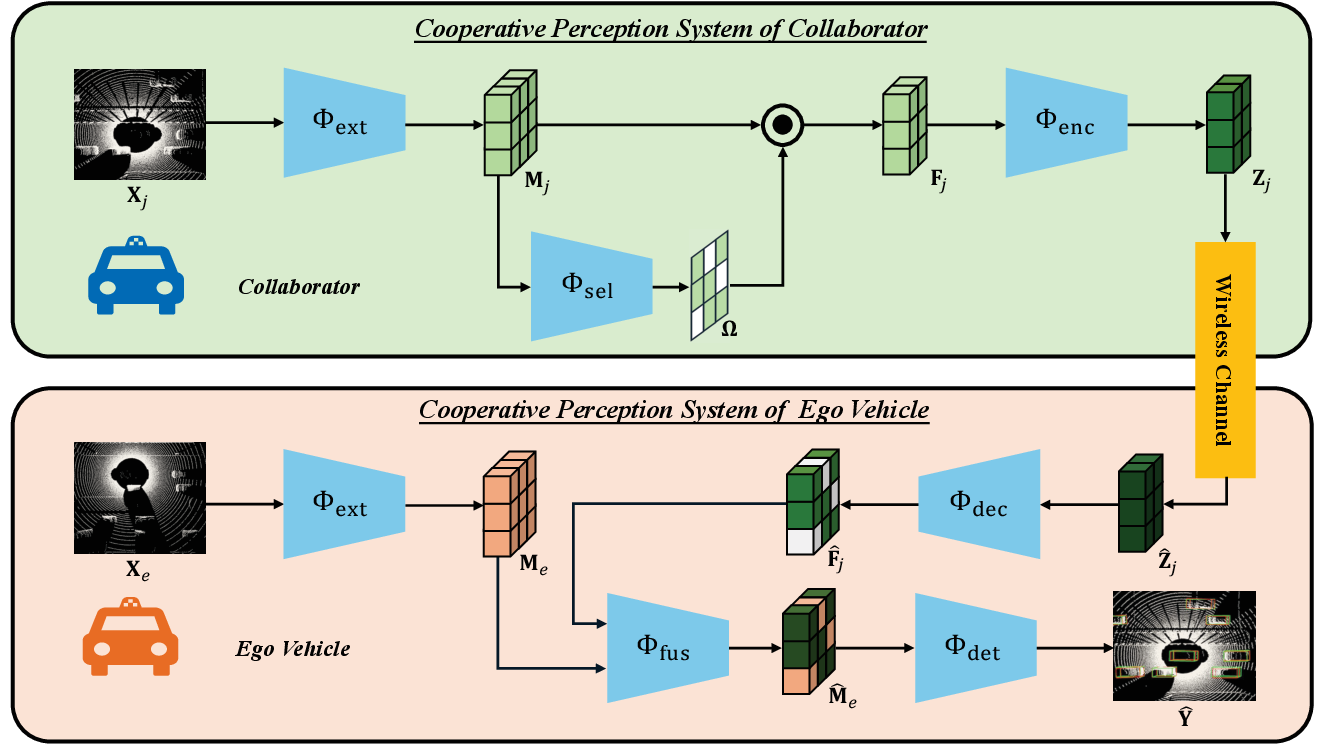}
  \caption{An illustration of the proposed SComCP framework.}
  \label{fig2}
\end{figure*}

\section{System Model and Problem Formulation}

In this section, we first introduce the system model of the cooperative perception system. Following this, we describe the primary objective of SComCP.
\subsection{System Model}

Consider a point-to-point communication scenario between a collaborating vehicle and an ego vehicle to enable an intermediate fusion-based collaborative perception, as illustrated in Fig.~\ref{fig2}. The perception task of interest is LiDAR-based 3D object detection. The collaborator employs LiDAR to perceive the surrounding road environment and transmits processed perception features to the ego vehicle via a noisy physical channel. Upon reception, the ego vehicle performs feature fusion by integrating the received perception features with its local features and utilizes the fused features for 3D object detection.
Specifically, this process involves the following five key steps.

\textbf{Feature extraction:} Let $\mathbf{X}_e$ and $\mathbf{X}_j$ represent the LiDAR point cloud data captured by the ego vehicle and the collaborator, respectively. To facilitate perception information sharing, the ego vehicle needs to broadcast its pose information to the collaborator. 
Assuming accurate vehicle poses and synchronized message transmissions, the collaborator projects its LiDAR point cloud data, $\mathbf{X}_j$, into the ego vehicle's coordinate frame upon receiving the ego vehicle's pose information, where the projection data is denoted by $\mathbf{X}_{j \rightarrow e}$. It is worth noting that such projection ensures that subsequent feature fusion is free from spatial misalignment errors.
Both the ego vehicle and the collaborator employ an identical feature extraction network, $\Phi_{\text{ext}}(\cdot)$, to convert the point cloud data into bird's-eye view (BEV) semantic feature maps, denoted by $\mathbf{M}_e \in \mathbb{R}^{H \times W \times C}$ and $\mathbf{M}_j \in \mathbb{R}^{H \times W \times C}$, respectively, where $H$, $W$, and $C$ represent the height, width, and the number of channels of the semantic feature map, respectively. The feature extraction process for the ego vehicle and the collaborator can be
\begin{equation}
    \label{eq1}
    {\mathbf{M}_e = \Phi_{\text{ext}}(\mathbf{X}_e)},
  \end{equation}
  and
  \begin{equation}
    \label{eq2}
    {\mathbf{M}_j = \Phi_{\text{ext}}(\mathbf{X}_{j \rightarrow e})}.
  \end{equation}

\textbf{Feature selection:} 
Due to the limited bandwidth of wireless networks and the stringent perception delay requirements imposed by autonomous driving, it may be quite challenging to transmit the entire semantic feature map. 
To achieve a better trade-off between perception performance and communication overhead, we adopt an importance-aware feature selection network, denoted as $\Phi_{\text{sel}}(\cdot)$, to identify the locations of semantic features most critical to the downstream perception task. Specifically, the collaborator generates a binary mask matrix, $\boldsymbol{\Omega} \in \mathbb{R}^{H \times W}$, based on the semantic feature map $\mathbf{M}_j$ using $\Phi_{\text{sel}}(\cdot)$, represented as
\begin{equation}
    \label{eq3}
    {\boldsymbol{\Omega} = \Phi_{\text{sel}}(\mathbf{M}_j)}.
  \end{equation}

\textcolor{black}{
The mask matrix $\boldsymbol{\Omega}$ is replicated $C$ times along the channel dimension and applied to the semantic feature map to filter out the semantic features with high semantic value, denoted by
\begin{equation}
\label{eq4}
\mathbf{F}_j = f_{\text{rep}}(\boldsymbol{\Omega}) \odot \mathbf{M}_j,
\end{equation}
where $\mathbf{F}_j \in \mathbb{R}^{K \times C}$ represents the selected semantic features, with $K$ denoting the number of semantic features selected from the semantic feature map across the $H \times W$ spatial dimensions, $f_{\text{rep}}(\cdot)$ denotes the replication function, and $\odot$ represents element-wise multiplication.
}

\textbf{Feature sharing:} The collaborator needs to transmit the selected semantic features $\mathbf{F}_j$ to the ego vehicle for fusion and further detection. However, in real-world wireless environments, signals are often distorted due to wireless channel conditions, leading to a significant drop in perception performance. To mitigate this, the collaborator employs a semantic encoder $\Phi_{\text{enc}}(\cdot)$ to encode the semantic features to the channel input symbols, represented as 
\begin{equation}
    \label{eq5}
    {\mathbf{Z}_j = \Phi_{\text{enc}}(\mathbf{F}_j)},
  \end{equation}
where \( \mathbf{Z}_j \in \mathbb{C}^{K \times C} \) are the channel input symbols, with \( K \times C \) denoting the number of symbols to be transmitted. 

After the encoding process, the collaborator transmits the channel input symbols $\mathbf{Z}_j$ to the ego vehicle via a wireless channel. 
Therefore, the channel output symbols $\mathbf{\hat{{Z}}}_j \in \mathbb{C}^{K \times C}$ received by the ego vehicle can be modeled as

\begin{equation}
    \label{eq6}
    {\mathbf{\hat{{Z}}}_j = \mathbf{h} \odot \mathbf{\mathbf{Z}}_j + \mathbf{n}},
  \end{equation}
where $\mathbf{h} \in \mathbb{C}^{K \times C}$ denotes the channel gain coefficient and $\mathbf{n} \in \mathbb{C}^{K \times C}$ represents the additive white Gaussian noise (AWGN). In particular, $\mathbf{n}$ consists of independent and identically distributed (i.i.d.) samples with the distribution $\mathcal{CN}(0, \sigma^2 )$, where $\sigma^2$ is the noise variance and $\mathcal{CN}(\cdot, \cdot)$ denotes a circularly symmetric complex Gaussian distribution.

At the receiver side, the ego vehicle decodes the received channel output symbols $\mathbf{\hat{{Z}}}_j$ through a semantic decoder $\Phi_{\text{dec}}(\cdot)$ to obtain the reconstructed semantic features $\mathbf{\hat{{F}}}_j \in \mathbb{R}^{K \times C}$, that is
\begin{equation}
    \label{eq7}
    {\mathbf{\hat{{F}}}_j = \Phi_{\text{dec}}(\mathbf{\hat{{Z}}}_j)}. 
  \end{equation}
  
\textbf{Feature fusion:} Through a fusion network, the ego vehicle enhances its semantic feature map by aggregating the semantic features received from the collaborator, enabling a more comprehensive perspective and scene understanding. The feature fusion process can be written as

\begin{equation}
    \label{eq8}
    {\mathbf{\hat{{M}}}_e = \Phi_{\text{fus}}(\mathbf{M}_e, \mathbf{\hat{{F}}}_j)},
  \end{equation}
where $\Phi_{\text{fus}}(\cdot)$ denotes the fusion network and $\mathbf{\hat{{M}}}_e \in \mathbb{R}^{H \times W \times C}$ is the fused semantic feature map integrating the semantic information from the collaborator and that from the ego vehicle.

\textbf{Object Detection:} After obtaining the fused semantic feature maps, a detection network $\Phi_{\text{det}}(\cdot)$ is employed to generate the detection result $\mathbf{\hat{{Y}}}$ for box regression and classification tasks, formulated as
\begin{equation}
    \label{eq9}
    {\mathbf{\hat{{Y}}} = \Phi_{\text{det}}(\mathbf{\hat{{M}}}_e)},
  \end{equation}
  where $\mathbf{\hat{{Y}}}$ comprises the regression and classification outputs. The regression output includes seven parameters $(x, y, z, w, l, h, \theta)$, including the center position $(x, y, z)$, the size $(w, l, h)$, and the yaw angle of the predefined bounding boxes $\theta$, enabling precise bounding box adjustment.  On the other hand, the classification output is the confidence score assigned to each bounding box, indicating the probability of an object existing in the bounding box. By combining the regression and classification outputs, the model achieves both category recognition and object localization.

\subsection{Problem Description}

The performance of our proposed ScomCP framework is evaluated in terms of average precision (AP), which is a widely used metric for assessing 3D object detection performance \cite{ap1}. AP quantifies detection accuracy by computing the area under the precision-recall curve, where a higher AP value indicates more accurate and reliable object detection. Let $\text{AP@}\text{IoU}_{\text{thr}} $ denote the AP value at the specific intersection-over-union (IoU) threshold $\text{IoU}_{\text{thr}}$, which can be represented as 
\begin{equation}
  \label{eq10}
  {\text{AP@}\text{IoU}_{\text{thr}} = f_{\text{eva}}(\mathbf{\hat{{Y}}}, \mathbf{Y}, \text{IoU}_{\text{thr}})},
\end{equation}
where $f_{\text{eva}}(\cdot)$ denotes the standard AP calculation function, and $\mathbf{Y}$ represents the ground truth of the object detection results \cite{ap}. Typically, a higher threshold means more stringent IoU requirement and thus leads to a lower AP value with the same detection result.

\textcolor{black}{The primary objective of ScomCP is to achieve an optimal trade-off between perception performance and communication overhead in real-world wireless networks.} In practical deployment scenarios, accounting for channel impairments is crucial for achieving the optimal balance. Therefore, our work focuses on the design and optimization of three core networks: importance-aware feature selection network $\Phi_{\text{sel}}(\cdot)$, semantic encoder $\Phi_{\text{enc}}(\cdot)$, and semantic decoder $\Phi_{\text{dec}}(\cdot)$. Specifically, $\Phi_{\text{sel}}(\cdot)$ is designed to select high-quality semantic features for perception tasks, while $\Phi_{\text{enc}}(\cdot)$ and $\Phi_{\text{dec}}(\cdot)$ are responsible for robust semantic feature encoding and decoding, respectively. 
We adhere to established standard network architectures for the remaining networks, including $\Phi_{\text{ext}}(\cdot)$, $\Phi_{\text{fus}}(\cdot)$, and $\Phi_{\text{det}}(\cdot)$, enabling us to concentrate our efforts on the crucial components, i.e., $\Phi_{\text{sel}}(\cdot)$,  $\Phi_{\text{enc}}(\cdot)$, and $\Phi_{\text{dec}}(\cdot)$.
With all modules well trained, the proposed end-to-end SComCP framework achieves a desirable trade-off between perception performance and communication overhead under dynamic channel conditions.

\section{Proposed Framework}

This section provides a detailed description of the proposed SComCP framework. First, we present an overview of the framework. Next, we describe the design of the importance-aware feature selection network and the semantic codec network in detail. Finally, we elaborate on the loss function design and the training strategy employed in the proposed framework.

\subsection{Framework Overview}
 The SComCP framework comprises five key components: feature extraction network, importance-aware feature selection network, semantic codec network, fusion network, and detection network, which collectively facilitate the entire collaborative perception process.

\textbf{Feature extraction network:} Given that the raw data captured by LiDAR contains rich spatial information but is highly redundant and unstructured, the feature extraction network is required to transform the high-dimensional raw input into compact and informative representations. In this work, we adopt a pillar feature network and a backbone module from the PointPillars architecture as the feature extraction network $\Phi_{\text{ext}}(\cdot)$ \cite{pointpillars}. Specifically, the raw point cloud data is first transformed into a stacked pillar tensor. 
Then, a scatter operation generates a 2D pseudo-image, which is subsequently passed through the backbone module for feature extraction.
This design effectively enables the conversion of raw LiDAR data into a BEV semantic feature map.

\textbf{Importance-aware feature selection network:}  The importance-aware feature selection network $\Phi_{\text{sel}}(\cdot)$ aims to distill high-value semantic features, $\mathbf{F}_j$, from the semantic feature map, $\mathbf{M}_j$, on the collaborator side. The detailed architecture of this module is discussed in the following subsection.

\textbf{Semantic codec network:}  
The semantic codec network comprises the semantic encoder $\Phi_{\text{enc}}(\cdot)$ and the semantic decoder $\Phi_{\text{dec}}(\cdot)$. $\Phi_{\text{enc}}(\cdot)$ maps semantic features into the channel input symbol representations for transmission. Conversely, $\Phi_{\text{dec}}(\cdot)$ reconstructs the semantic features from the distorted channel output symbols received under imperfect channel conditions. The architectural details are presented in the next subsection.

\textbf{Fusion network:} Existing collaborative perception studies primarily employ attention-based feature fusion strategies \cite{where2comm, vit}. In our framework, the fusion network $\Phi_{\text{fus}}(\cdot)$ utilizes a scaled dot-product attention mechanism to achieve cross-view feature integration \cite{opencood}. By projecting the collaborator’s point cloud into the ego vehicle’s coordinate frame, the network effectively aggregates features from the collaborator based on the spatial correlations between the aligned ego vehicle and the collaborator. This spatially-aware fusion mechanism efficiently leverages spatial context while maintaining lower computational overhead compared to global attention approaches \cite{scaled}.

\textbf{Detection network:}  
The detection network $\Phi_{\text{det}}(\cdot)$ consists of two independent branches, corresponding to the box regression and classification tasks, with each branch implemented using a $1 \times 1$ convolutional layer \cite{opencood}.

\begin{figure}[t!]
  \centering
  \includegraphics[width=3.5in]{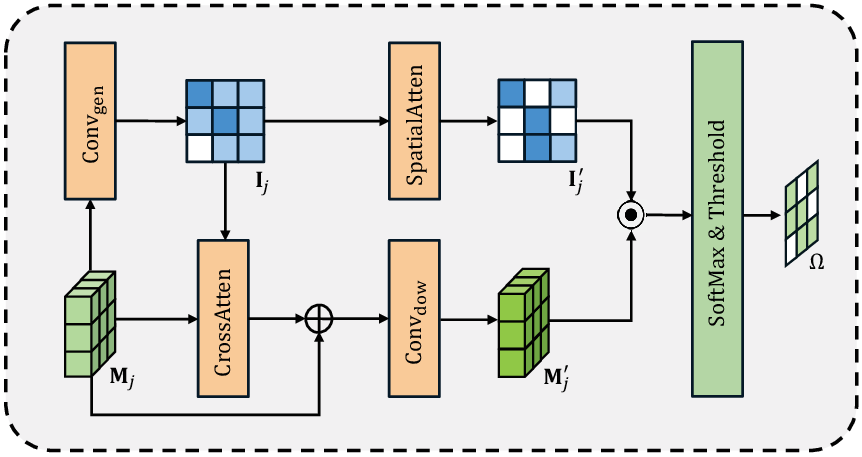}
  \caption{Network structure of our proposed importance-aware feature selection network.}
  \label{fig3}
\end{figure}

\subsection{Importance-Aware Feature Selection Network}

In collaborative perception scenarios, the semantic feature map $\mathbf{M}_j$ typically contains significant redundancy, where many features lack sufficient semantic value or provide marginal perceptual value to the ego vehicle. Transmitting the entire semantic feature map incurs substantial communication overhead without commensurate gains in perception performance. To address this issue, the importance-aware feature selection network is proposed to dynamically select high-value semantic features that are critical to downstream detection tasks. Furthermore, the network adaptively adjusts the number of semantic features to be transmitted based on the semantic richness of the current scene, thereby further enhancing transmission efficiency.

The architecture of the importance-aware feature selection network is illustrated in Fig.~\ref{fig3}. Specifically, the semantic feature map $\mathbf{M}_j$ is first passed through an importance map generator, $\text{Conv}_{\text{gen}}(\cdot)$, to generate the importance map $\mathbf{I}_j \in \mathbb{R}^{H \times W}$, given by
\begin{equation}
    \label{eq11}
    {\mathbf{I}_j = \text{Conv}_{\text{gen}}(\mathbf{M}_j)},
  \end{equation}
where each pixel value in $\mathbf{I}_j$ indicates the confidence that the corresponding region contains an object. The module $\text{Conv}_{\text{gen}}(\cdot)$ is implemented as the classification branch of the detection network \cite{where2comm}.

Next, a cross-attention mechanism $\text{CrossAtten}(\cdot)$ is employed to effectively associate the importance map $\mathbf{I}_j$ with the semantic feature map $\mathbf{M}_j$, where $\mathbf{I}_j$ serves as the query, while $\mathbf{M}_j$ acts as both the key and the value \cite{attention}. To align the channel dimensions with the importance map, a downsampling convolutional layer $\text{Conv}_{\text{dow}}(\cdot)$ is applied. Then, the enhanced semantic feature map $\mathbf{M}_j'$ can be represented as
\begin{equation}
    \label{eq12}
    {\mathbf{M}_j' = \text{Conv}_{\text{dow}} \left( \mathbf{M}_j + \gamma \cdot \text{CrossAtten}(\mathbf{I}_j, \mathbf{M}_j) \right)},
  \end{equation}
where $\gamma$ is a learnable parameter that controls the residual fusion of the attention-enhanced feature map with the original input, enabling more accurate feature refinement.

\begin{figure*}[htbp]
  \centering
  \includegraphics[width=6.2in]{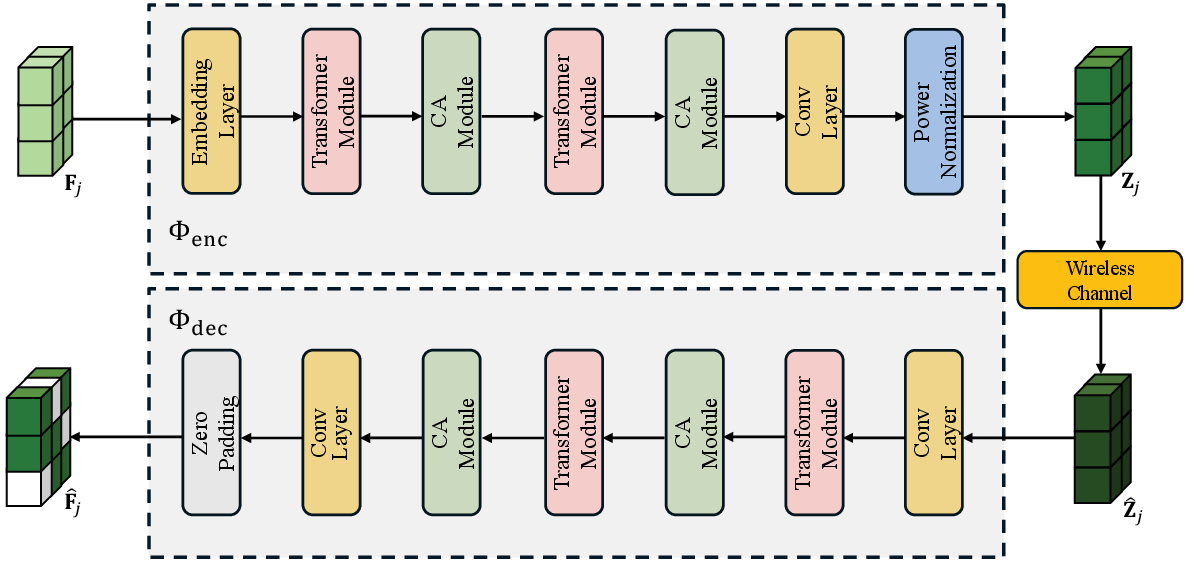}
  \caption{Network structure of our proposed semantic codec network.}
  \label{fig4}
\end{figure*}

On the other hand, to enhance the saliency of potential object regions in the spatial dimension, a spatial attention module, denoted as $\text{SpatialAtten}(\cdot)$, is applied to the importance map \cite{cbam} to obtain the spatial attention refined importance map $\mathbf{I}_j'$, represented to 
\begin{equation}
    \label{eq13}
    {\mathbf{I}_j' = \text{SpatialAtten}(\mathbf{I}_j)}.
  \end{equation}

Subsequently, element-wise multiplication between $\mathbf{M}_j'$ and $\mathbf{I}_j'$ is performed to enhance the object regions while suppressing background noise. The resulting vector is fed into the soft-max operator, $\text{SoftMax}(\cdot)$, yielding the probability of each feature in the semantic feature map, $\mathbf{P}_{\text{keep}}$. The probability is proportional to the importance of the corresponding feature and is given by

\begin{equation}
    \label{eq14}
    {\mathbf{P}_{\text{keep}} = \text{SoftMax}(\mathbf{M}_j' \odot \mathbf{I}_j')}.
  \end{equation}

Furthermore, by utilizing $\mathbf{P}_{\text{keep}}$ and a threshold $\gamma$, we can dynamically select the most semantically significant features for transmission. Specifically, the binary mask matrix $\boldsymbol{\Omega}$ is computed as
\begin{equation}
    \label{eq15}
    {\boldsymbol{\Omega} = I(\mathbf{P}_{\text{keep}} > \gamma)},
  \end{equation}
where $I(\cdot)$ is an indicator function. 
In this binary mask, entries with a value of 1 correspond to retained features, while entries with a value of 0 indicate features deemed unimportant and thus discarded.
By adjusting the threshold $\gamma$, the number of retained semantic features can be controlled, enabling content-adaptive variable length coding. Specifically, a lower value of $\gamma$ increases the feature retention rate, thereby leading to a higher transmission load.
 
In summary, this importance-aware feature selection network not only effectively reduces communication overhead but also provides high-quality perception information for subsequent feature fusion.

\subsection{Semantic Codec Network}

This subsection introduces the proposed semantic codec network, the architecture of which is shown in Fig.~\ref{fig4}. Specifically, the proposed semantic codec consists of a semantic encoder $\Phi_{\text{enc}}(\cdot) $ at the transmitter and a semantic decoder $ \Phi_{\text{dec}}(\cdot) $ at the receiver. Both components primarily include channel-attention (CA) modules and transformer modules \cite{attention}. A detailed description is provided below.

\begin{figure}[t!]
  \centering
  \includegraphics[width=3.5in]{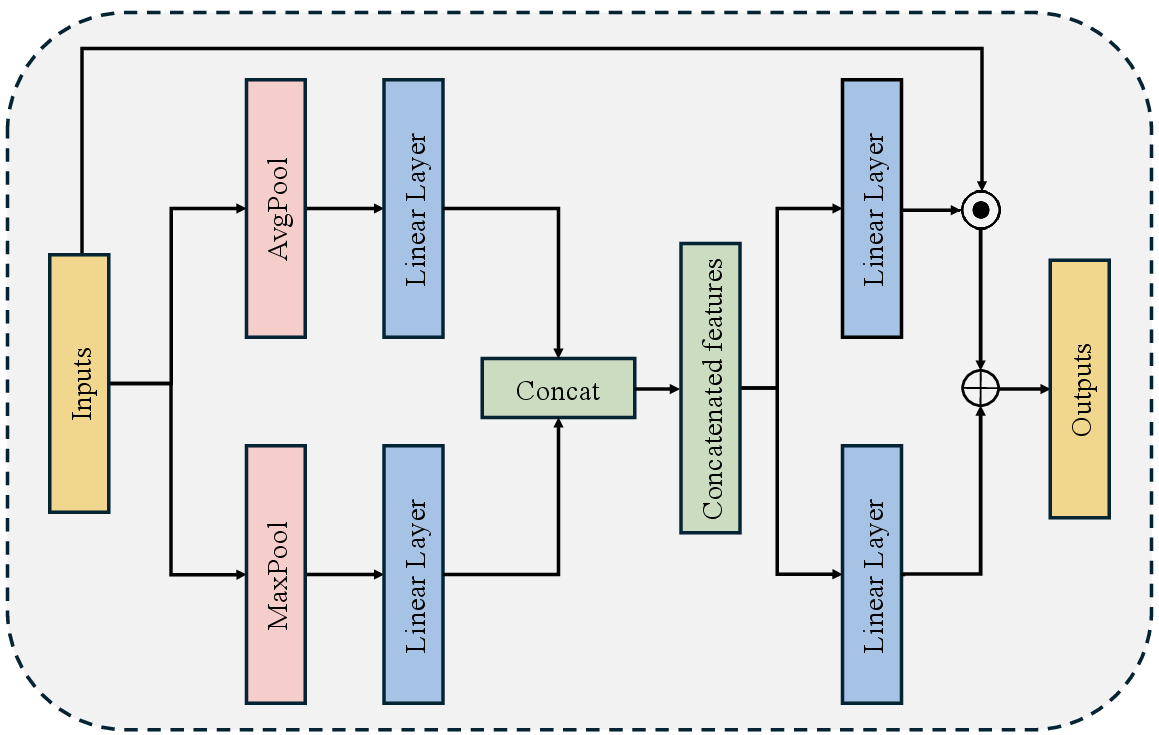}
  \caption{Network structure of our proposed CA module.}
  \label{fig5}
\end{figure}

To refine the semantic features along the channel dimension, we propose the CA module that dynamically recalibrates channel-wise feature responses. Following the approach in \cite{cbam}, the CA module is designed as shown in Fig.~\ref{fig5}. Specifically, taking the $k$-th CA module as an example, it first extracts global context from the input features, $\mathbf{f}_k$, via parallel average pooling operation, $\text{AvgPool}(\cdot)$, and max pooling operation, $\text{MaxPool}(\cdot)$. After that, we employ a linear layer, Linear(.), to further process the extracted global context, giving rise to
\begin{equation}
  \label{eq17}
  { \mathbf{s}_k^{avg} =  \text{Linear}(\text{AvgPool}(\mathbf{f}_k))},
\end{equation}
and
\begin{equation}
  \label{eq18}
  { \mathbf{s}_k^{max} =  \text{Linear}(\text{MaxPool}(\mathbf{f}_k))},
\end{equation}
where $\mathbf{s}_k^{avg}$ and $\mathbf{s}_k^{max}$ are the final average-pooled and max-pooled features. 
Subsequently, the $\mathbf{s}_k^{avg}$ and $\mathbf{s}_k^{max}$ are concatenated to generate the concatenated features $\mathbf{w}_k$, given by
\begin{equation}
  \label{eq19}
  { \mathbf{w}_k = \text{Concat}(\mathbf{s}_k^{avg}, \mathbf{s}_k^{max})},
\end{equation}
where $\text{Concat}(\cdot, \cdot)$ denotes the concatenation operation. The concatenated features are then passed through two independent linear layers to generate scaling weights and an addition offset. The scaling weights dynamically scale the input features to emphasize task-critical channels, while the addition offset adjusts the offset of the scaled features to refine the feature representation. Finally, the module outputs the sum of the scaled and offset-adjusted features, ensuring that the weight-modified semantic features are better adapted for channel transmission. Accordingly, the output feature of the $k$-th CA module can be derived as
\begin{equation}
  \label{eq21}
  {\mathbf{f}_{k+1} = \text{Linear}(\mathbf{w}_k) \cdot \mathbf{f}_k + \text{Linear}(\mathbf{w}_k)}.
\end{equation}

The semantic encoder $\Phi_{\text{enc}}(\cdot)$ unifies source encoding, channel encoding, and modulation within a single neural architecture for JSCC. Through end-to-end training, $\Phi_{\text{enc}}(\cdot)$ can effectively transform the semantic features into an interference-resistant channel input symbol representation suitable for transmission over noisy physical channels, thus enabling semantic-level information transmission with enhanced efficiency and noise resilience. Specifically, in the semantic encoder, the input semantic features $\mathbf{F}_j$ are first mapped to a higher dimensional space via an embedding layer to extract richer feature information. Then, transformer modules and CA modules work in an alternating connection manner. The transformer modules learn effective features of the input sequences by modeling long-range dependencies, while the CA modules generate scaling parameters based on the output of the transformer module. These parameters selectively enhance the most relevant feature channels, thus optimizing the feature representation of the transformer module. Functionally, the output of the CA module can be seen as a filtered and optimized version of the output from the previous layer (transformer module). This alternating processing mechanism allows the network to capture both the global dependencies and the local details of the input features. Finally, the semantic encoder maps the features to the complex space through a convolution layer and performs power normalization to ensure that the output signal's power meets the preset constraint conditions. Specifically, the channel input symbols $\mathbf{Z}_j$ are normalized to satisfy the average power constraint, as given by
\begin{equation}
  \label{eq17}
  { P_z = \frac{E[\mathbf{Z}_j^* \mathbf{Z}_j]}{K \cdot C} \leq P_{\text{bound}}},
\end{equation}
where $\mathbf{Z}_j^*$ denotes the conjugate transpose of $\mathbf{Z}_j$, $P_z$ represents the average power of $\mathbf{Z}_j$, and  $P_{\text{bound}}$ is the average power constraint, which is typically set to $1$ to maintain generality.

At the receiver side, the semantic decoder mirrors the encoder architecture, integrating source decoding, channel decoding, and demodulation. The decoder employs the same components as the encoder, including the transformer module and the CA module, to ensure structural symmetry. Additionally, at the output of the semantic decoder, zero-padding is applied to restore the semantic features to their original dimensions, ensuring consistency with the dimensions of $\mathbf{M}_j$.

Overall, the proposed semantic codec network enhances the representation of transmitted semantic features, thereby improving the robustness of SComCP against channel impairments.

\subsection{Loss Function and Training Strategy}
 
The following presents the design of the loss function used to train the SComCP framework. As outlined in Algorithm 1, the SComCP framework employs a three-stage training strategy in which the importance-aware feature selection network and the semantic codec network are trained separately before the entire network is optimized jointly.
It is important to note that during the first two stages, the remaining modules—including the feature extraction network, feature fusion network, and detection network—utilize pre-trained models from open-source repositories. These components remain frozen and do not participate in the training process during these initial stages.

In the first stage, the importance-aware feature selection network is trained. During this stage, the training data sequentially passes through the feature extraction network, feature selection network, fusion network, and detection network. Given that the output of each network serves as the input to the subsequent one, an end-to-end training approach is employed. Since the semantic codec network has not yet been trained, the impact of the wireless channel is not considered, assuming lossless data transmission. The loss function for this stage is designed for the 3D object detection task, aiming to improve detection accuracy. Therefore, we follow the PointPillars method for the loss function design, which jointly considers the classification loss and the regression loss \cite{pointpillars}. In particular, the classification loss is computed using focal loss \cite{focalloss}, while the regression loss is evaluated using smooth L1 loss. The overall loss function for the first stage is given as
\begin{equation}
  \label{eq22}
  {\mathcal{L}_{\text{per}} = \mathcal{L}_{\text{cls}}(\mathbf{\hat{Y}}, \mathbf{Y}) + \eta \cdot \mathcal{L}_{\text{reg}}(\mathbf{\hat{Y}}, \mathbf{Y})},
\end{equation}
where $\mathcal{L}_{\text{cls}}(\cdot)$, $\mathcal{L}_{\text{reg}}(\cdot)$, and $\eta$ represent the classification loss, the regression loss, and the weight factor, respectively.

In the second stage, the semantic codec network is trained under the Rayleigh fading channel, where all networks remain frozen. The noisy physical channel is integrated as a non-trainable layer within the semantic codec network, enabling end-to-end training.
The loss function in this stage additionally accounts for the transmission loss, which aims to recover the semantic features disrupted by channel impairments. This transmission loss is specifically designed to restore the semantic features that have been degraded during transmission and is quantified by the mean squared error (MSE) between the semantic encoder input and the semantic decoder output.
Therefore, the overall loss function for this stage takes into account both the object detection loss and the transmission loss, as given by
\begin{equation}
  \label{eq23}
  {\mathcal{L}_{\text{trans}} = \mathcal{L}_{\text{cls}}(\mathbf{\hat{{Y}}}, \mathbf{Y}) + \eta \cdot \mathcal{L}_{\text{reg}}(\mathbf{\hat{{Y}}}, \mathbf{Y}) + \gamma \cdot \mathcal{L}_{\text{mse}}(\mathbf{\hat{{F}}}_j, \mathbf{F}_j)},
\end{equation}
where $\mathcal{L}_{\text{mse}}(\cdot)$ and $\gamma$ represent the MSE loss, and its corresponding weight factor, respectively. 

In the third stage, the complete SComCP framework undergoes end-to-end training under the Rayleigh fading channel, with all constituent networks trained simultaneously. The loss function employed in this stage is identical to that used in the first stage, $\mathcal{L}_{\text{per}}$, and is designed to optimize the overall system performance.

\begin{algorithm}[ht]
\caption{Training Algorithm}
\label{alg1}
\KwIn{LiDAR point cloud data $\mathbf{X}_e$ and $\mathbf{X}_j$}
\KwOut{Detection result $\hat{\mathbf{Y}}$}
\vspace{1mm}
\textbf{First stage:} Train the importance-aware feature selection network.

\While{not converged}{
  \For{ $\mathbf{X}_e$ and $\mathbf{X}_j$ in batch samples}{
     
    \vspace{0.2mm}
    $\mathbf{M}_e = \Phi_{\text{ext}}(\mathbf{X}_e)$,
    $\mathbf{M}_j = \Phi_{\text{ext}}(\mathbf{X}_{j \rightarrow e})$\;
    \vspace{0.2mm}
    $\boldsymbol{\Omega} = \Phi_{\text{sel}}(\mathbf{M}_j)$,
    $\mathbf{F}_j = f_{\text{rep}}(\boldsymbol{\Omega}) \odot \mathbf{M}_j$\;
    \vspace{0.2mm}
    $\hat{\mathbf{M}}_e = \Phi_{\text{fuse}}(\mathbf{M}_e, \mathbf{F}_j)$\;
    \vspace{0.2mm}
    $\hat{\mathbf{Y}} = \Phi_{\text{det}}(\hat{\mathbf{M}}_e)$\;
 
  }
  Compute loss $\mathcal{L}_{\text{per}}$ and update $\Phi_{\text{sel}}(\cdot)$ via gradient descent.
}

\vspace{1mm}
\textbf{Second stage:} Train the semantic codec network.

\While{not converged}{
  \For{ $\mathbf{X}_e$ and $\mathbf{X}_j$ in batch samples}{
     
    \vspace{0.2mm}
    $\mathbf{M}_e = \Phi_{\text{ext}}(\mathbf{X}_e)$,
    $\mathbf{M}_j = \Phi_{\text{ext}}(\mathbf{X}_{j \rightarrow e})$\;
    \vspace{0.2mm}
    $\boldsymbol{\Omega} = \Phi_{\text{sel}}(\mathbf{M}_j)$,
    $\mathbf{F}_j = f_{\text{rep}}(\boldsymbol{\Omega}) \odot \mathbf{M}_j$\;
    \vspace{0.2mm}
    $\mathbf{Z}_j = \Phi_{\text{enc}}(\mathbf{F}_j)$\;
    \vspace{0.2mm}
    $\hat{\mathbf{Z}}_j = \mathbf{h} \odot \mathbf{Z}_j + \mathbf{n}$\;
    \vspace{0.2mm}
    $\hat{\mathbf{F}}_j = \Phi_{\text{dec}}(\hat{\mathbf{Z}}_j)$\;
    \vspace{0.2mm}
    $\hat{\mathbf{M}}_e = \Phi_{\text{fuse}}(\mathbf{M}_e, \hat{\mathbf{F}}_j)$\;
    \vspace{0.2mm}
    $\hat{\mathbf{Y}} = \Phi_{\text{det}}(\hat{\mathbf{M}}_e)$\;
 
  }
  Compute loss $\mathcal{L}_{\text{trans}}$ and update $\Phi_{\text{enc}}(\cdot), \Phi_{\text{dec}}(\cdot)$ via gradient descent.

}

\vspace{1mm}
\textbf{Third stage:} Train the whole network.

\While{not converged}{
  \For{ $\mathbf{X}_e$ and $\mathbf{X}_j$ in batch samples}{
     \vspace{0.2mm}
  
    $\mathbf{M}_e = \Phi_{\text{ext}}(\mathbf{X}_e)$,
    $\mathbf{M}_j = \Phi_{\text{ext}}(\mathbf{X}_{j \rightarrow e})$\;
    \vspace{0.2mm}
    $\boldsymbol{\Omega} = \Phi_{\text{sel}}(\mathbf{M}_j)$,
    $\mathbf{F}_j = f_{\text{rep}}(\boldsymbol{\Omega}) \odot \mathbf{M}_j$\;

    \vspace{0.2mm}
 
    $\mathbf{Z}_j = \Phi_{\text{enc}}(\mathbf{F}_j)$\;
    \vspace{0.2mm}
    $\hat{\mathbf{Z}}_j = \mathbf{h} \odot \mathbf{Z}_j + \mathbf{n}$\;
    \vspace{0.2mm}
    $\hat{\mathbf{F}}_j = \Phi_{\text{dec}}(\hat{\mathbf{Z}}_j)$\;

    \vspace{0.2mm}
 
    $\hat{\mathbf{M}}_e = \Phi_{\text{fuse}}(\mathbf{M}_e, \hat{\mathbf{F}}_j)$\;
    \vspace{0.2mm}
    $\hat{\mathbf{Y}} = \Phi_{\text{det}}(\hat{\mathbf{M}}_e)$\;
 
  }
  Compute loss $\mathcal{L}_{\text{per}}$ and update the whole network via gradient descent.

}
\end{algorithm}

\section{Simultation Results}	
This section provides a comprehensive performance evaluation of our proposed ScomCP framework. 
\subsection{Experimental Setup}
\textbf{Dataset.}
 We evaluated all methods on the OPV2V dataset which is a large-scale open dataset for V2V collaborative perception \cite{opencood}. This dataset was generated using the OpenCDA platform, which integrates the CARLA and SUMO simulators \cite{carla}. It comprises 12K frames of synchronized LiDAR point clouds and RGB images, with 230K annotated 3D bounding boxes. Following the default configuration, the dataset is divided into 6,764 frames for training, 1,981 frames for validation, and 2,170 frames for testing. Moreover, the maximum collaboration distance between the ego vehicle and collaborators is set to 70 meters. Object detection is evaluated within the range $x \in [-140, 140]\ \text{m}$ and $y \in [-40, 40]\ \text{m}$, with the ego vehicle coordinate system as the reference. All experiments and comparative analyses are conducted under this standardized setup.

\textbf{Implementation Details.}
The proposed model is implemented in PyTorch and trained on an NVIDIA GeForce RTX 4080 GPU. We employ the Adam optimizer for backpropagation, with an initial learning rate of $1 \times 10^{-3}$, and an exponential weight decay of $0.6$. The batch size is set to $4$. To simulate practical wireless environments, each training sample is assigned an SNR value uniformly sampled from the range of $0$ dB to $20$ dB. Additionally, the proposed model is trained only under Rayleigh fading channels, while performance evaluation is conducted under both AWGN and Rayleigh fading channels.

\begin{figure*}[t]
  \centering
  \begin{minipage}[b]{0.49\textwidth}
      \centering
      \includegraphics[width=\linewidth]{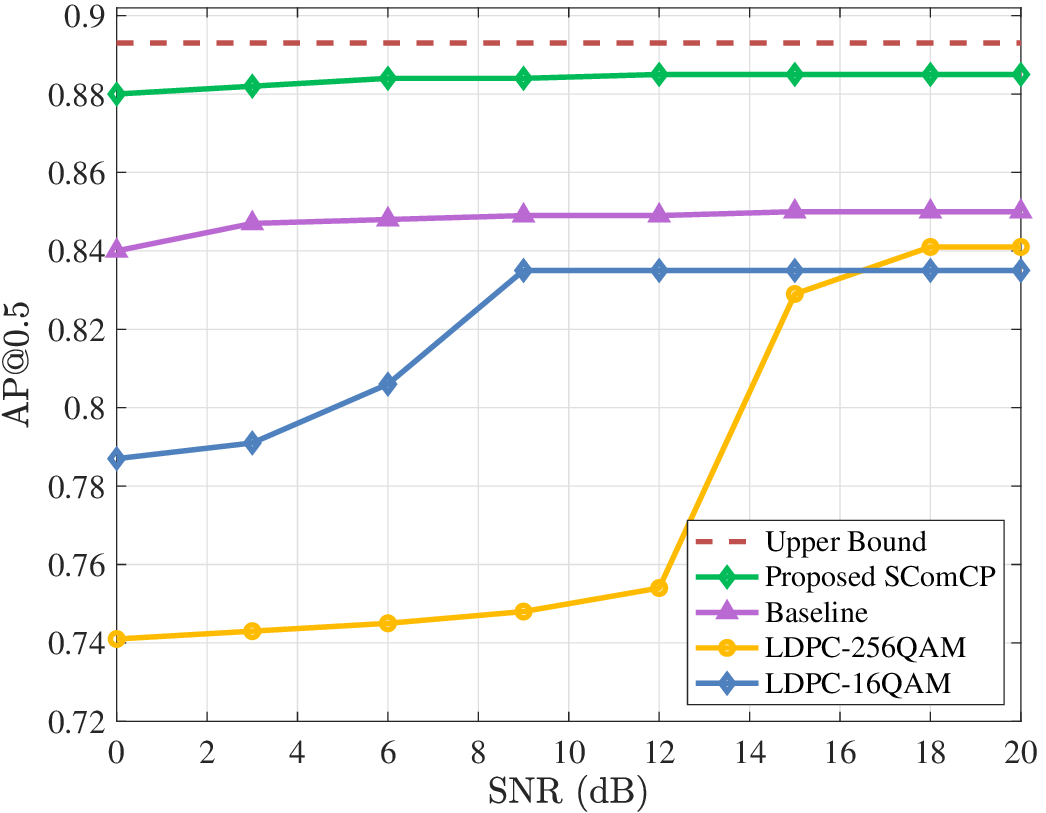}
      \centerline{(a) Detection accuracy in $\text{AP@}0.5$.}
  \end{minipage}
  \hspace{0.005\textwidth}
  \begin{minipage}[b]{0.49\textwidth}
      \centering
      \includegraphics[width=\linewidth]{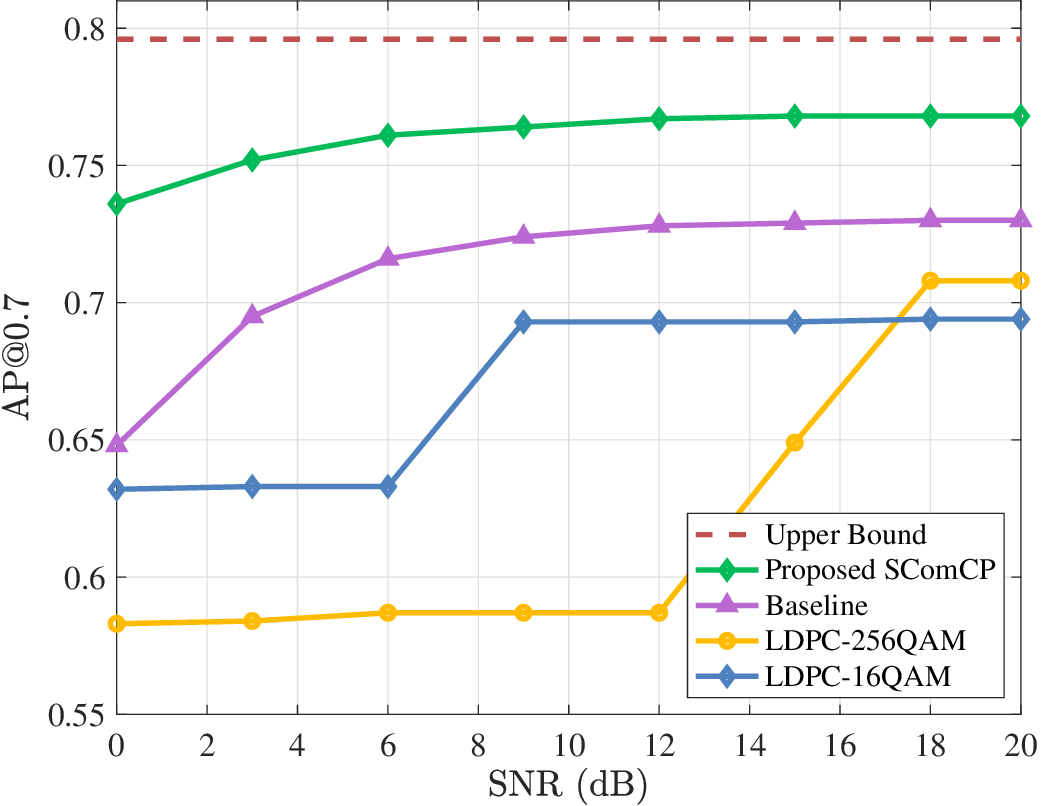}
      \centerline{(b) Detection accuracy in $\text{AP@}0.7$.}
  \end{minipage}
  \caption{Performance of our proposed method compared with other schemes at different SNRs over an AWGN channel.
  }
  \label{fig6}
\end{figure*}

\textbf{Comparison Schemes.}
We compare the SComCP scheme with the following alternatives:

  •	Baseline Scheme: This scheme is currently a state-of-the-art method \cite{franklin}. It leverages the importance map from Where2Comm to reduce bandwidth consumption and integrates a CNN-based semantic codec network to mitigate channel impairments \cite{where2comm}.

  •	Traditional Scheme: The method also employs the importance map to reduce the transmitted data volume. Subsequently, uniform 8-bit quantization is applied for source coding of the selected features.
  For channel coding, a 1/2 rate LDPC code with a codeword length of 1000 bits is utilized, followed by  16QAM or 256QAM modulation.

	•	Upper Bound: The complete semantic feature map generated by the backbone network are transmitted, with the physical channel assumed to be ideal. This represents a theoretical upper bound on the performance of the intermediate fusion scheme.

  To ensure a fair comparison, all models use PointPillars as the backbone network.

\subsection{Evaluation Metrics}

We adopt $\text{AP@}0.5$ and $\text{AP@}0.7$ as the primary evaluation metrics. Additionally, during the transmission process, the SNR significantly impacts the quality of the reconstructed signal, thereby affecting the final perception performance. 
 The SNR is defined as the ratio of the average power $P_z$ of the channel input symbols  $\mathbf{Z}_j$ to the average noise power, expressed in dB form as
\begin{equation}
  \label{eq24}
  {\text{SNR(dB)} = 10 \log_{10} \left( \frac{P_z}{\sigma^2} \right)}.
\end{equation}

The compression ratio (CR) is defined as the proportion of non-zero elements in the mask matrix $\boldsymbol{\Omega}$, which represents the effective size of the transmitted semantic features. However, under the same CR, the traditional scheme incurs varying channel uses due to differences in LDPC code rates and modulation orders (e.g., 16QAM vs. 256QAM). To ensure a fair comparison, the CR is adjusted to equalize the number of channel uses across all schemes. The total number of channel uses is calculated as
\begin{equation}
  \label{eq25}
  \text{Channel Uses} = \frac{S_m \times \mathrm{CR} \times 8}{R_c \times \log_2 M_c},
\end{equation}
where $S_m$ is the size of the semantic feature map $\mathbf{M}_j$, $M_c$ is the order of the modulation,  and $R_c$ is the rate of the LDPC code.
Based on (25), the CR values for the 1/2 rate LDPC code with a 16QAM modulation and 1/2 rate LDPC code with a 256QAM modulation are set to $\text{CR} = 3.5\times10^{-4}$ and $\text{CR} = 7\times10^{-4}$, respectively, ensuring the same channel usage as the proposed method with an average CR of $\text{CR} = 1.4\times10^{-3}$.

\subsection{Performance Analysis}

\begin{figure*}[h]
  \centering
  \begin{minipage}[b]{0.49\textwidth}
      \centering
      \includegraphics[width=\linewidth]{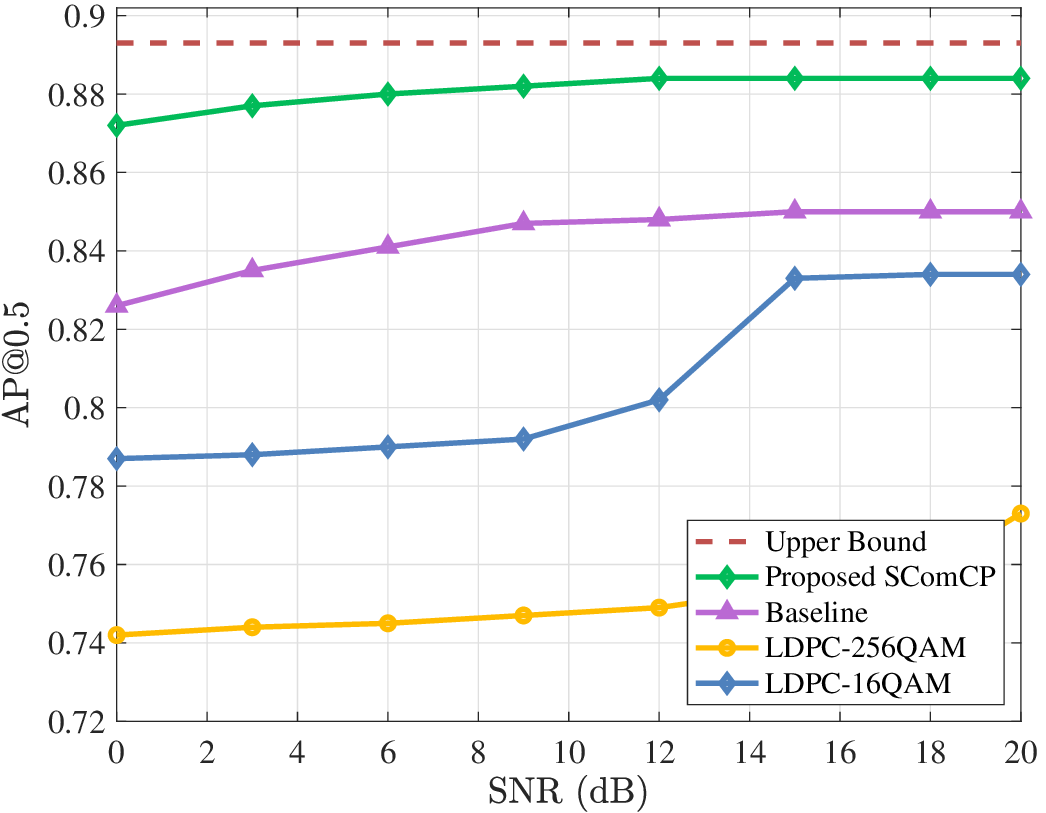}
      \centerline{(a) Detection accuracy in $\text{AP@}0.5$.}
  \end{minipage}
  \hspace{0.005\textwidth}
  \begin{minipage}[b]{0.49\textwidth}
      \centering
      \includegraphics[width=\linewidth]{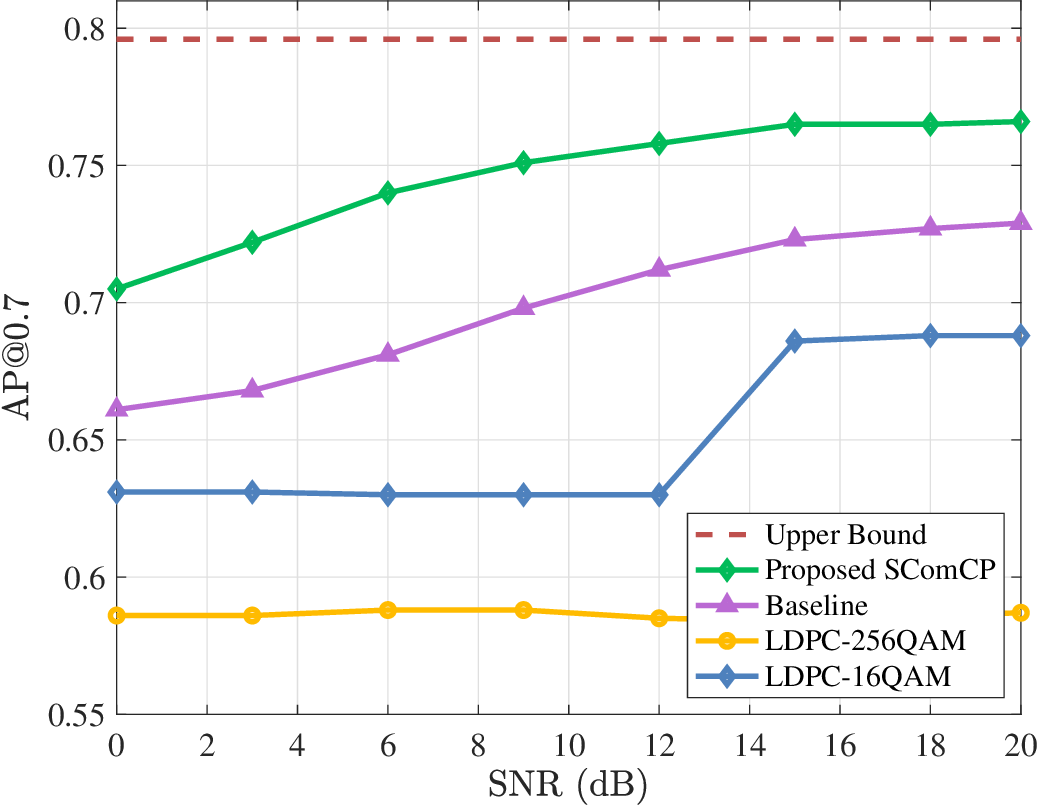}
      \centerline{(b) Detection accuracy in $\text{AP@}0.7$.}
  \end{minipage}
  \caption{Performance of our proposed method compared with other schemes at different SNRs over a Rayleigh channel.}
  \label{fig7}
\end{figure*}

Fig.~\ref{fig6} illustrates a comparison of perception performance between the proposed SComCP and other methods under varying SNRs in an AWGN channel. As shown in Fig.~\ref{fig6}(a), the proposed method outperforms other schemes in terms of $\text{AP@}0.5$. 
As expected, the upper bound achieves the highest perception performance since it transmits all perception features under the assumption of lossless transmission. However, its high communication overhead and reliance on idealized assumptions render it impractical for real-world wireless communication scenarios. In contrast, our method strikes an excellent balance, achieving superior perception-communication trade-offs compared to all comparison schemes. In a scenario of $\text{CR} = 1.4 \times10^{-3}$ and $\text{SNR} = 0$ dB, the performance of SComCP is only $1.3\%$ lower than the upper bound, demonstrating remarkable perception-communication efficiency. Moreover, compared to the baseline scheme, SComCP shows a $4.0\%$ improvement at $\text{SNR} = 0$ dB, validating the effectiveness of the proposed SComCP framework. 
In the traditional scheme, 256QAM outperforms 16QAM in high SNR regimes, as it can carry more information. However, higher-order modulation is more sensitive to noise due to the closer spacing of its constellation points, resulting in lower performance than 16QAM under low SNR conditions. Meanwhile, both schemes suffer from the “cliff effect'' in low SNR regimes due to the inherent limitations of traditional communication systems.
In contrast, our proposed deep JSCC-based SComCP framework is not subject to the “cliff effect''. Unlike traditional scheme, which experience abrupt performance degradation in low SNR regimes, SComCP exhibits a gradual and stable decline in performance as channel conditions worsen. This gain comes from the task-oriented semantic communication design, which prioritizes the transmission of semantic information over precise bit-level accuracy.

As shown in Fig.~\ref{fig6}(b), we compare the performance of the proposed method with other schemes in terms of $\text{AP@}0.7$. The experimental results, similar to those for $\text{AP@}0.5$, demonstrate that our method significantly outperforms other schemes, particularly under low SNR conditions. 
It can be observed that the performance of all methods at $\text{AP@}0.7$ drops significantly compared to $\text{AP@}0.5$. This is due to the more stringent detection criterion imposed by a higher IoU threshold, which requires more precise object recognition and localization, leading to lower AP scores. In terms of $\text{AP@}0.7$, the performance gap between the proposed method and the upper bound becomes more pronounced. This is because $\text{AP@}0.7$ is more sensitive to noise, and even slight distortions in semantic features can lead to significant performance degradation. In contrast, the upper bound assumes ideal lossless transmission, providing error-free perceptual information, thereby resulting in better performance.
Additionally, the performance improvement of our proposed SComCP over the baseline scheme increases to $8.8\%$ at $0$ dB. This is because SComCP is more robust in low SNR scenarios, providing higher-quality perceptual information. While this advantage is less pronounced at $\text{AP@}0.5$, the stricter detection criteria of $\text{AP@}0.7$ more clearly highlight the robustness of SComCP in low SNR scenarios. By avoiding catastrophic performance losses in such conditions, SComCP ensures robust perception capabilities, which are critical for safety-sensitive autonomous driving applications.

\textcolor{black}{
Fig.\ref{fig7} compares the performance of our proposed method with other schemes in terms of $\text{AP@}0.5$ and $\text{AP@}0.7$ at different SNR levels over a Rayleigh channel. While all methods exhibit performance degradation under Rayleigh channel conditions compared to the AWGN channel, our proposed approach consistently outperforms the alternatives across different SNR levels, which is consistent with the results shown in Fig.\ref{fig6}. Notably, our model was trained exclusively under Rayleigh channel conditions but demonstrates excellent performance when tested on both Rayleigh and AWGN channels without requiring retraining. This robustness across varying channel conditions highlights the practical feasibility of our approach for real-world deployments, where channel characteristics may fluctuate.}

\begin{figure*}[t]
  \centering
  \includegraphics[width=6.5in]{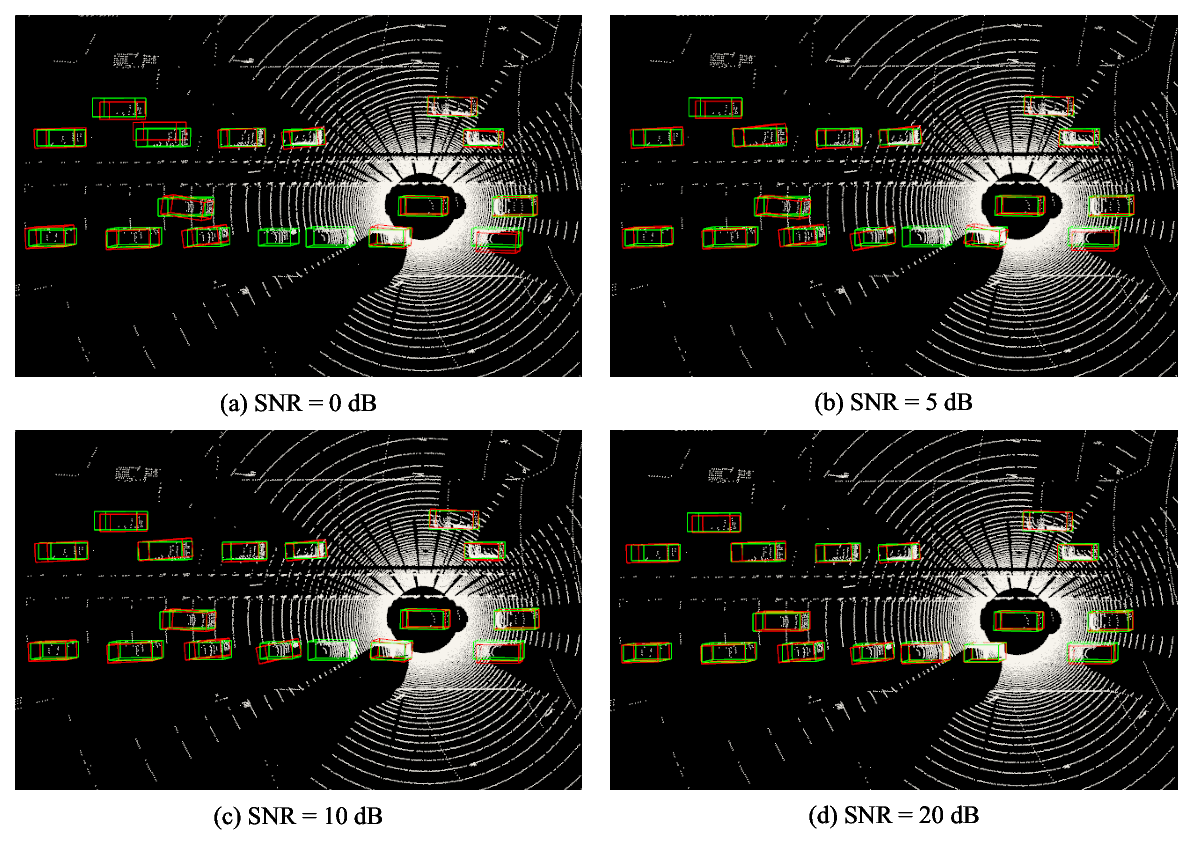}
  \caption{Examples of 3D detection outputs produced by our proposed method.}
  \label{fig8}
\end{figure*}

Fig.~\ref{fig8} illustrates a collaborative perception case where the ego vehicle detects surrounding objects from collaborators. The visualization shows predicted detections (red boxes) and ground truth (green boxes) under different SNR scenarios over an AWGN channel. In low SNR scenarios, as shown in Fig.~\ref{fig8}(a) and~\ref{fig8}(b), corruption of transmitted semantic features leads to incomplete object detection, with several vehicles remaining undetected. Conversely, in high SNR scenarios, as shown in Fig.~\ref{fig8}(c) and~\ref{fig8}(d), the high overlap between predicted detections and ground truth boxes confirms the reliable performance of our method. These results validate the effectiveness of our approach across varying SNR scenarios.

\subsection{Ablation Study}

\begin{figure}[htbp]
  \centering
  \includegraphics[width=3.5in]{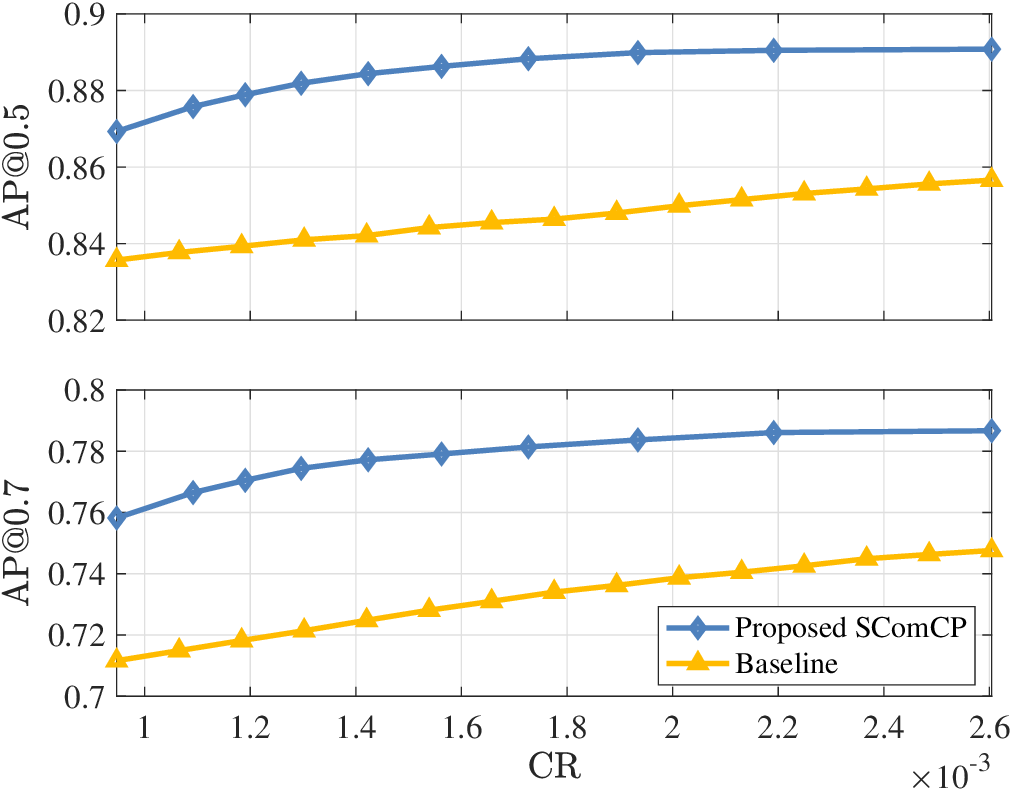}
  \caption{Performance of our proposed importance-aware feature selection network compared with the baseline scheme at different values of CR in the perfect transmission scenario.}
  \label{fig9}
\end{figure}

\begin{figure}[htbp]
  \centering
  \includegraphics[width=3.5in]{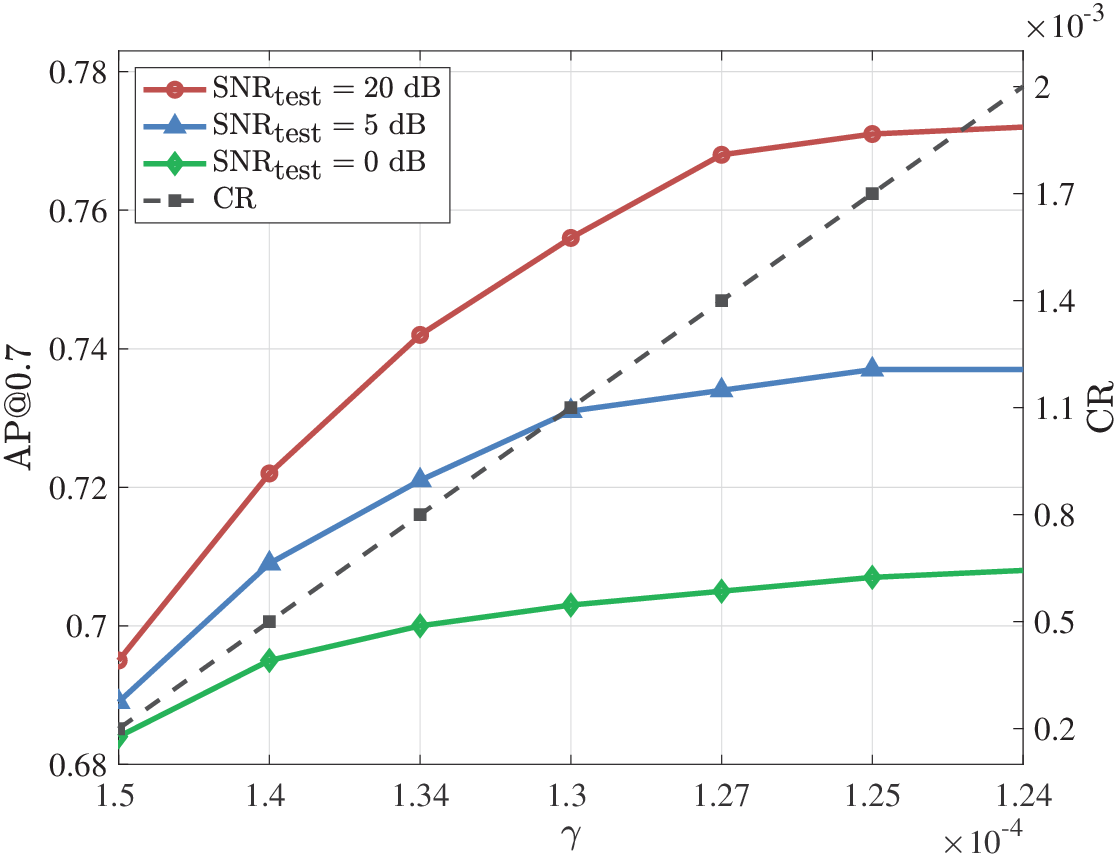}
  \caption{Performance of our proposed method with different thresholds over a Rayleigh channel evaluated in terms of $\text{AP@}0.7$.}
  \label{fig10}
\end{figure}

In this subsection, we present the results of an ablation study to independently evaluate the impact of the importance-aware feature selection network and the semantic codec network.

Fig.~\ref{fig9} compares the performance of our proposed SComCP with the baseline scheme under lossless transmission without semantic codec networks. Specifically, the blue and yellow curves represent the performance evaluation of the importance-aware feature selection network and Where2Comm under varying CRs, respectively, demonstrating the trade-off between perception performance and communication overhead. We observe the following: (i) In terms of $\text{AP@}0.5$ and $\text{AP@}0.7$, SComCP consistently outperforms the baseline scheme at the same CR, validating the effectiveness of the importance-aware feature selection network; (ii) As CR increases, the perception performance of both SComCP and the baseline scheme improves, as sharing more semantic features between vehicles provides additional information for better detection; (iii) As CR increases, the perception performance gain of SComCP diminishes and eventually saturates at $\text{CR} = 2.2 \times10^{-3}$, while the baseline scheme continues to improve. This is because the proposed importance-aware feature selection network dynamically adjusts the amount of transmitted semantic features based on the semantic richness of the scene, leading to better transmission efficiency compared to the baseline scheme. 
\textcolor{black}{
Complementing these results, Fig.~\ref{fig10} illustrates the performance of SComCP over a Rayleigh channel at different values of $\gamma$, evaluated in terms of $\text{AP@}0.7$. As a key parameter, $\gamma$ directly influences the CR by controlling the selection threshold for high-value semantic features. The results demonstrate that SComCP maintains consistently strong perception performance across all tested CR and SNR scenarios, highlighting its robustness. Moreover, the flexibility introduced by $\gamma$ enables our approach to adapt dynamically to the semantic richness of different scenes. In semantically rich environments, the model selectively transmits a greater volume of critical information to preserve perception quality, while in simpler scenes, transmission resources are conserved without compromising performance. This adaptive behavior not only enhances overall transmission efficiency but also ensures reliable perception across diverse communication environments.}

\begin{figure}[htbp]
  \centering
  \includegraphics[width=3.5in]{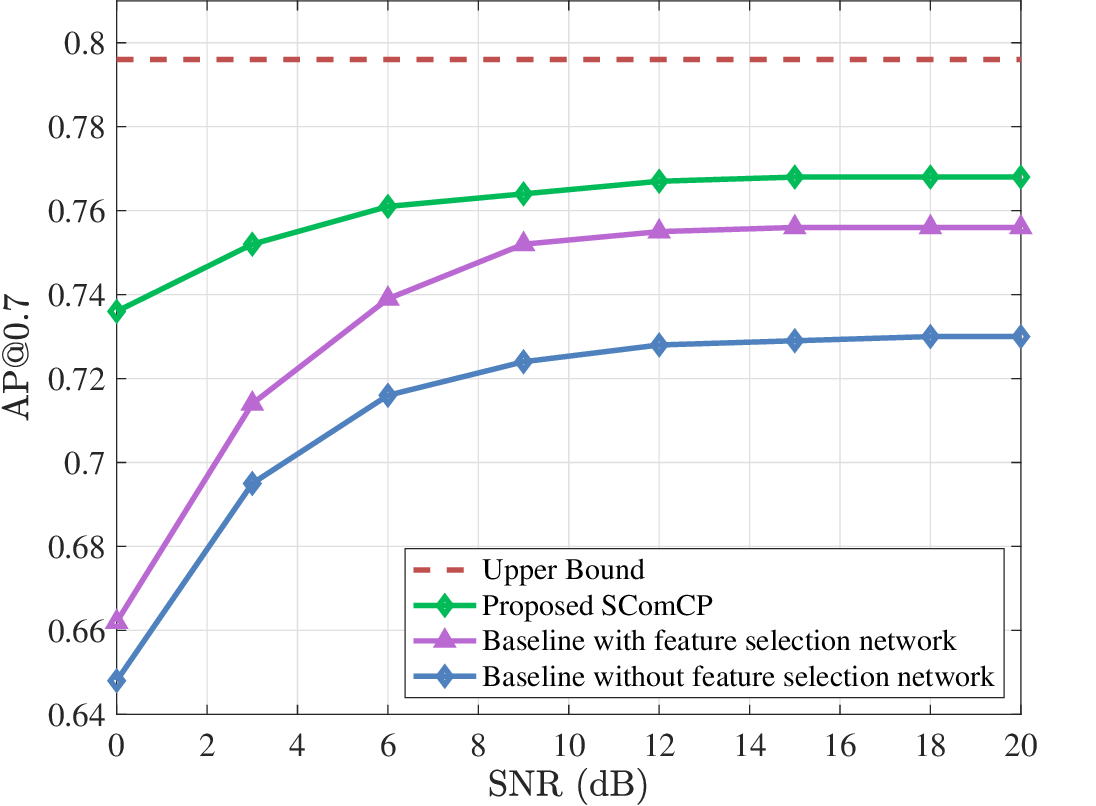}
  \caption{Performance of our proposed method compared with other schemes at different SNRs over an AWGN channel evaluated in terms of $\text{AP@}0.7$.}
  \label{fig11}
\end{figure}

\begin{figure}[htbp]
  \centering
  \includegraphics[width=3.5in]{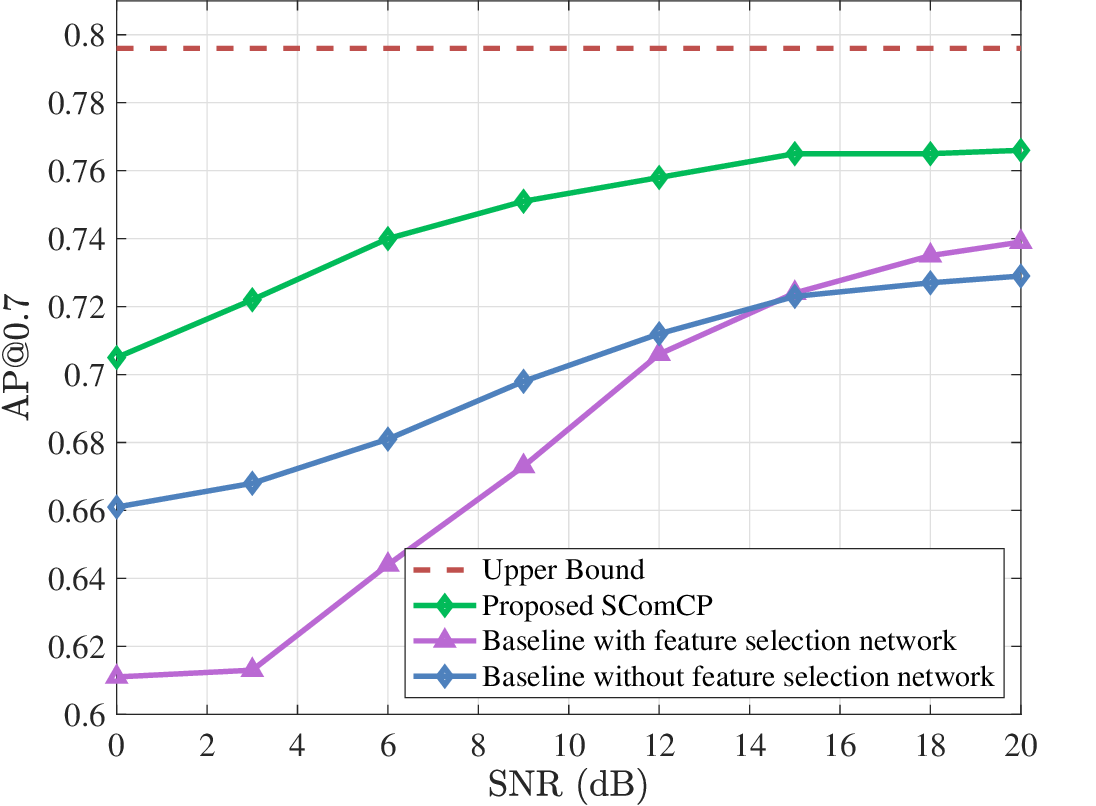}
  \caption{Performance of our proposed method compared with other schemes at different SNRs over a Rayleigh channel evaluated in terms of $\text{AP@}0.7$.}
  \label{fig12}
\end{figure}

\textcolor{black}{
Fig.~\ref{fig11} and Fig.~\ref{fig12} compare the performance of our proposed semantic codec network with other methods under different SNRs over both AWGN and Rayleigh channels. To ensure a fair comparison of semantic codec network performance, the baseline scheme also utilizes the importance-aware feature selection network, which has been shown to effectively select high-value semantic features while maintaining the same number of channel uses.
As illustrated in Fig.~\ref{fig11}, the baseline scheme with the feature selection network outperforms the baseline scheme without it, attributable to the effectiveness of the importance-aware feature selection network. However, despite this enhancement, a performance gap persists between the enhanced baseline and our proposed SComCP. Specifically, at $\text{SNR} = 20$ dB, SComCP maintains a $1.2\%$ advantage over the baseline scheme with the feature selection network. This advantage becomes increasingly pronounced as channel conditions worsen, expanding to $7.3\%$ at $\text{SNR} = 0$ dB. These results highlight the advantages of the proposed semantic codec network, particularly under challenging low-SNR conditions.}

As shown in Fig.~\ref{fig12}, it is observed that the baseline scheme performs worse in low SNR scenarios when the importance-aware feature selection network is used, compared to when it is not. This can be attributed to a key limitation that neither the CNN-based encoder network in the baseline scheme nor the conventional traditional scheme effectively mitigates channel impairments. While the proposed importance-aware feature selection network selects high-quality semantic features, channel impairments severely disrupt these features during transmission. The subsequent fusion of these distorted semantic features with the ego vehicle's semantic feature maps propagates errors, leading to a $4.9\%$ decrease in perception performance at $\text{SNR} = 0$ dB compared to the case without the feature selection network. In contrast, prior approaches like the baseline scheme, where the selected features are of moderate quality, are less susceptible to channel impairments. However, the high-quality features extracted by our network, while beneficial under clean channel conditions, are particularly susceptible to corruption in noisy environments. The proposed SComCP addresses this challenge effectively through its robust semantic codec network design, demonstrating significant resilience to channel impairments, especially in low SNR scenarios.


\section{Conclusion}	

This paper presented SComCP, a task-oriented semantic communication framework for V2V collaborative perception. The framework was designed to optimize the trade-off between perception performance and communication overhead in realistic communication environments. Specifically, we proposed an importance-aware feature selection network that identified and prioritized high-value semantic features, thereby reducing communication overhead while enhancing perception performance. Complementing this, we developed a deep JSCC-based semantic codec network that effectively mitigates channel impairments, thereby enhancing the robustness of data transmission, particularly in low-SNR scenarios. Experimental results demonstrated that the proposed SComCP framework achieved a superior balance between perception and communication efficiency.

\bibliographystyle{IEEEtran}
\bibliography{IEEEabrv,reference}

\vfill
\end{document}